\documentclass[showpacs,prl,floatfix,twocolumn,amsmath,a4]{revtex4}
\usepackage{graphicx}

\begin{document}

\title{Magnetically induced electronic ferroelectricity in half-doped manganites}

\author{Gianluca Giovannetti$^{1,2}$}
\author{Sanjeev Kumar$^{1,2}$}
\author{Jeroen van den Brink$^{1,3}$}
\author{Silvia Picozzi$^4$}
\affiliation{$^1$Institute Lorentz for Theoretical Physics, Leiden University, Leiden, The Netherlands}
\affiliation{$^2$Faculty of Science and Technology and MESA+ Research Institute,
University of Twente, The Netherlands}
\affiliation{$^3$Institute for Molecules and Materials, Radboud Universiteit, Nijmegen, The Netherlands}
\affiliation{$^4$Consiglio Nazionale delle Ricerche - Istituto Nazionale per la Fisica della Materia (CNR-INFM), CASTI Regional Laboratory, 67100 L'Aquila, Italy}
\date{\today}

\begin{abstract}
Using a joint approach of density functional theory and model calculations, we demonstrate that a prototypical charge ordered half-doped manganite, La$_{1/2}$Ca$_{1/2}$MnO$_3$ is multiferroic. The combination of a peculiar charge-orbital ordering and a tendency to form spin dimers breaks inversion symmetry, leads to a ferroelectric ground-state with a polarization up to several $\mu C/cm^2$. The presence of improper ferroelectricity does not depend on hotly debated structural details of this material: in the Zener-polaron structure we find a similar dramatic ferroelectric response with a large polarization of purely magnetic origin. 
\end{abstract}


\maketitle 

Materials with simultaneous magnetic and ferroelectric ordering --multiferroics-- are attracting enormous scientific interest~\cite{Cheong07,Ramesh07}. They offer the potential to control the magnetic orderparameter by the ferroelectric one and {\it vice versa} --a very desirable property from a technological point of view~\cite{Eerenstein06}. Such control requires large multiferroic couplings: a substantial ferroelectric polarization needs to be induced by the magnetic ordering. Even if in quite a few materials ferroelectricity and magnetism coexist, multiferroic couplings are tiny~\cite{Kimura03,Hur04,Giovannetti08,Lottermoser04}. When designing materials with large multiferroic couplings, one has to exclude from the start the largest class of multiferroics, the ones in which multiferroicity is driven by spiral magnetic ordering. In these materials multiferroicity relies on relativistic spin-orbit coupling as a driving force, which is intrinsically weak~\cite{Cheong07}. Charge ordered magnetic compounds are a far more promising class of materials with potentially large multiferroic couplings~\cite{Brink08}. Coexistence of charge ordering and magnetism is found in a substantial number of transition metal oxides~\cite{Imada98}, but to become strongly multiferroic a material needs to meet three additional requirements: $(i)$ the symmetry is such that the magnetic ordering can push the charge ordering pattern from site-centered to bond-centered or {\it vice versa}~\cite{Efremov04}, $(ii)$ the material is insulating, as it has to support a ferroelectric polarization and $(iii)$ the material is electronically soft, so that inside it charge can easily be displaced. Half doped manganites of the type La$_{1/2}$Ca$_{1/2}$MnO$_3$ famously meet the last two requirements~\cite{Milward05}. Here we show they also fulfill the first one, allowing strong multiferroic coupling to emerge.

In a combined approach of ab-initio density functional and model Hamiltonian calculations we show that the strong electron-electron interactions together with the Jahn-Teller lattice distortions that are present in this manganite cause a canting instability of its magnetic groundstate, driving a reconstruction of its charge ordering from site-centered towards bond-centered. The resulting non-collinear magnetic ordering induces in La$_{1/2}$Ca$_{1/2}$MnO$_3$ a purely electronic polarization of several $\mu$C/cm$^2$, almost two orders of magnitude larger than the one of a typical multiferroic such as TbMn$_2$O$_5$~\cite{Hur04}. 

\begin{figure}
\resizebox{80mm}{!}{\includegraphics{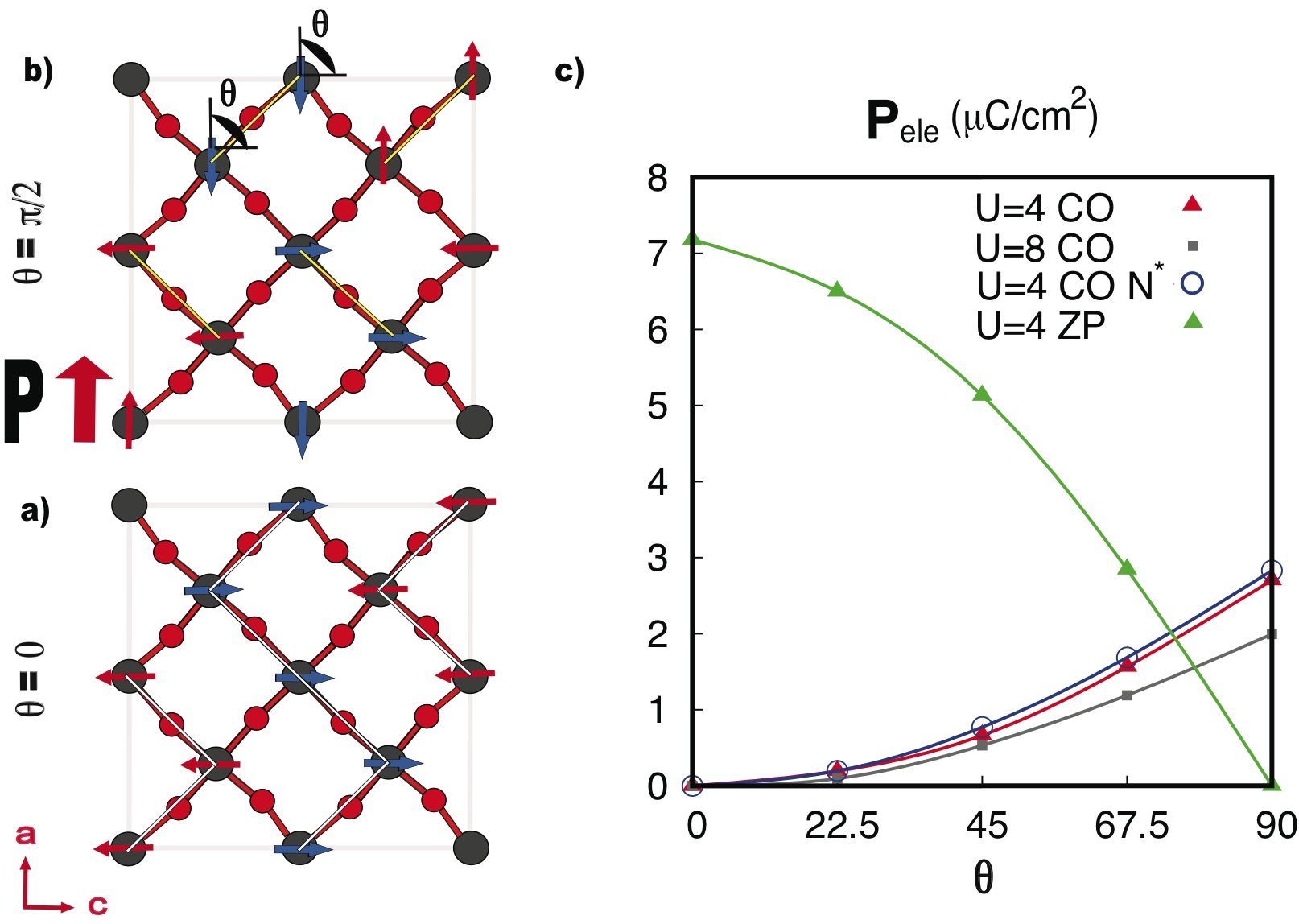}}
\caption{(Color online)
Spin-directions in the MnO$_2$ plane for a) $\theta$ =0$^o$ (AFM-CE) and b) $\theta$ = 90$^o$ ($\perp$). 
Zig-zag chains and FM dimers highlighted. c) Electronic polarization $P_{ele}$ (in $\mu
C/Cm^2$) vs $\theta$ (in degrees) for U = 4 eV and U = 8 eV for the
experimental ionic structure of Ref.~\cite{radaelli} (CO) and of Ref.\cite{rodriguez} (ZP). ``Half-doping" is simulated both with {\it i}) a checkerboard 
arrangement of La and Ca ions and {\it ii}) by adding half extra-electron for each Ca ion in  CaMnO$_3$ and compensating with a homogeneous positive background (denoted as N$^*$)} 
\label{fig:1}
\end{figure}

We consider the half-doped manganite, La$_{1/2}$Ca$_{1/2}$MnO$_3$ (LCMO) in the experimentally observed antiferromagnetic (AFM) CE double-zigzag spin state~\cite{radaelli,wollan}  (see Fig.\ref{fig:1}a), which is stable below $T_N \sim$ 155 K. Despite having a long history~\cite{wollan,goodenough}, the experimental crystallographic and corresponding electronic structure of LCMO  (as that of the closely related Pr$_{1-x}$Ca$_x$MnO$_3$, $x \sim$ 0.5) is still debated. On one hand, a traditional checkerboard  CO state has been proposed~\cite{radaelli},  given by the alternation of orbitally-ordered Mn$^{3+}$ (at the center of Jahn-Teller distorted octahedra) and Mn$^{4+}$ (in largely undistorted octahedra) with a mostly site-centered (SC) CO.  On the other hand, a bond-centered (BC) model has been suggested~\cite{daoud,wu}, based on the so-called Zener-polaron (ZP) state where equivalent Mn $d^4$ ions, showing no charge-disproportionation (CD), couple into ferromagnetic (FM) dimers sharing a spin-polarized hole on the intermediate O atom. In both SC and BC cases, we find magnetic groundstate structures that break inversion symmetry (IS) and result in polar states with relatively strong ferroelectricity~\cite{Betouras07}.

First-principles studies have proven to be tremendously helpful to shed light on the microscopic origin and on the quantitative evaluation of the electric polarization $P$ in several improper magnetic ferroelectrics~\cite{prlslv,prlgg,xe,whangbo}. Our density functional theory (DFT) simulations are performed within the generalized gradient approximation (GGA)~\cite{pbe} to the exchange-correlation potential and treating the Mn $d$ electrons via a Hubbard-like potential within GGA+U~\cite{anisimov} (unless otherwise noted, U = 4 eV, J = 0.9 eV). We used the Vienna Ab-initio Simulation Package (VASP)~\cite{vasp}, including the non-collinear-spin formalism~\cite{hobbs} and the Berry-phase (BP) approach to evaluate electric polarization P~\cite{berry}. The cut-off for the plane-wave basis set was chosen as 400 eV for the collinear and non-collinear spin-configurations and a [3,3,4]  mesh was used for the  Brillouin-zone sampling. In the BP approach, we integrated over 12 {\bf k}-space strings parallel to either the $a$ or $b$ axis, each string divided in 8 {\bf k}-points. The experimental lattice parameters and ionic positions were taken from Ref.~\cite{radaelli} for the CO-like structure ($P2_1/m$ space group) and from Ref.~\cite{rodriguez} for the ZP-like structure ($P2_1nm$ space group).

\begin{figure}
\resizebox{80mm}{!}{\includegraphics[angle=0]{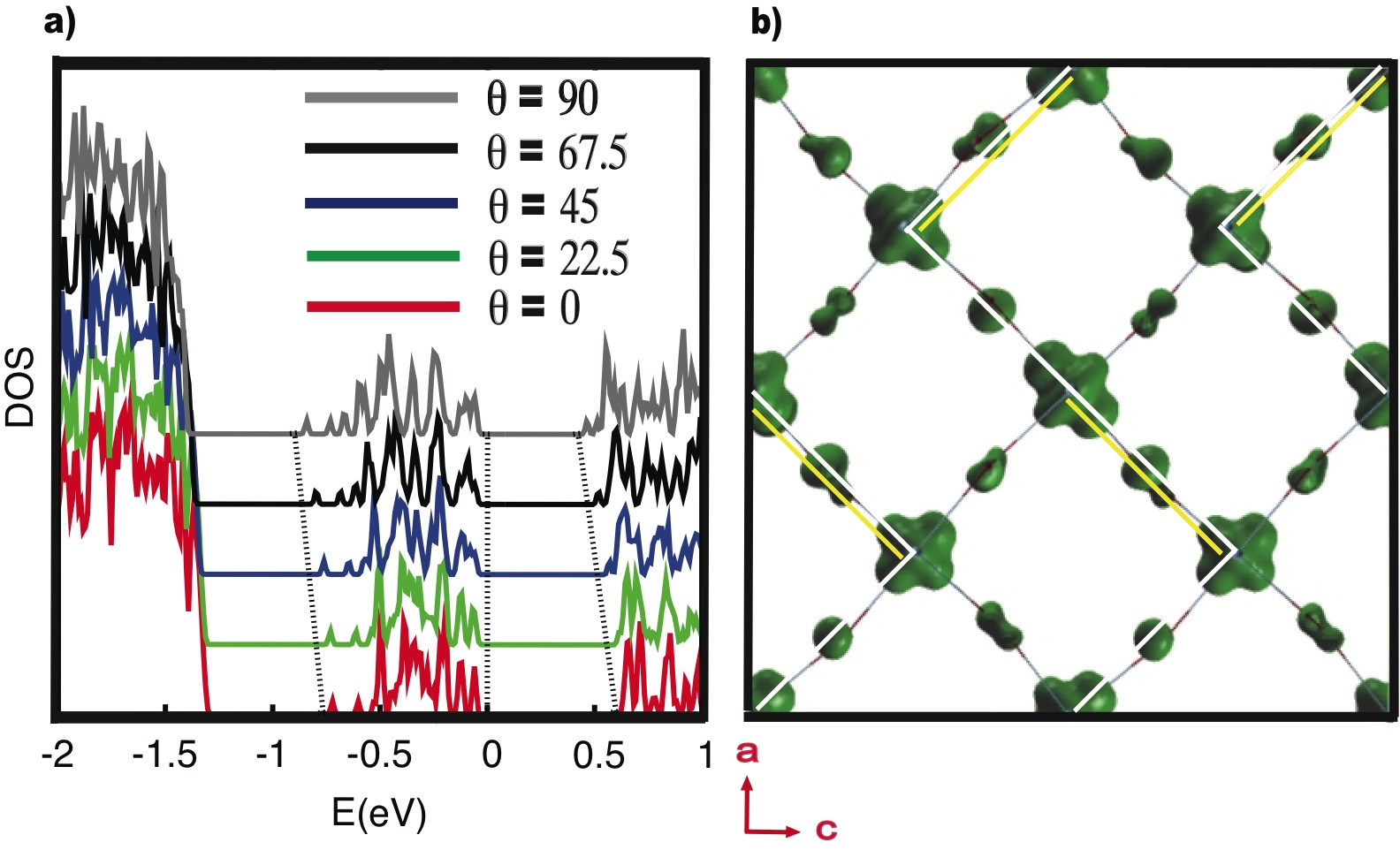}}
\caption{(Color online) DFT results:
a) Total DOS for different $\theta$: DOS are arbitrarily shifted on the $y$-axis with a shift proportional to $\theta$. The zero of the energy scale marks the Fermi level. b) Isosurface of the $e_g$ bands in the $\perp$ structure: view on the MnO$_2$ plane.} 
\label{fig:2}
\end{figure}

From our  calculations we find that the CE-type AFM (Fig.~\ref{fig:1}a) is insulating (see Fig.~\ref{fig:2}a) and clearly orbitally ordered, with a small CD, consistent with  previous collinear-spin electronic-structure works~\cite{satpathy}. For U = 4 eV, there is a clear gap, the CD amounts to $\delta \sim$ 0.15 $e^-$ whereas the magnetic moments are $\sim$ 3.3 $\mu_B$ and 3.05 $\mu_B$ on the nominally Mn$^{3+}$ and  Mn$^{4+}$, respectively.  The BP calculation of $P$ shows, as expected,  that the centrosymmetric CE-type structure is paraelectric.  However, a spin-rotation immediately induces a ferroelectric moment. We consider a rotation of the spins of two neighboring Mn (one 3+ and one 4+) along the up spin-chain by an angle $\theta$ and, correspondingly, one dimer in the down spin-chain by the same angle. This particular spin rotation is motivated by the fact that it tunes the magnetic CE state continuously towards the one compatible with ZP structure (denoted as $\perp$ in what follows), which corresponds to $\theta$ = 90$^o$ and is shown in Fig.~\ref{fig:1}b. The calculated electronic polarization $P_{ele}$ increases monotonically with $\theta$, reaching $\sim$ 3 $\mu C/cm^2$ for $\theta$ = 90$^o$. Importantly, the results are stable with respect to an increase in the value of $U$, and to the details of how the half doping is achieved within DFT (see Fig.~\ref{fig:1}c).

The resulting total density of states (DOS) shows that a larger value of $\theta$ broadens the $e_g$ band ($~$0.7 eV wide), therefore reducing the band-gap (see Fig.~\ref{fig:2}a). In Fig.~\ref{fig:2}b we plot an isosurface of the $e_g$-bands, indicating a clear orbital ordering (OO), with two kinds of  different Mn: the nominal Mn$^{3+}$ shows a staggered $(3x^2-r^2)/(3y^2-r^2)$ orbital arrangement, whereas the nominal Mn$^{4+}$  shows a much more isotropic charge-distribution, as given by partial occupation of both $(3z^2-r^2)$ and $(x^2-y^2)$ orbitals. Since this situation is by far similar to the collinear CE case (not shown), it proves that the spin-rotation alone does not alter significantly the CO/OO. 

\begin{figure}
\resizebox{80mm}{!}{\includegraphics[width = 8cm, angle=0 , clip = true]{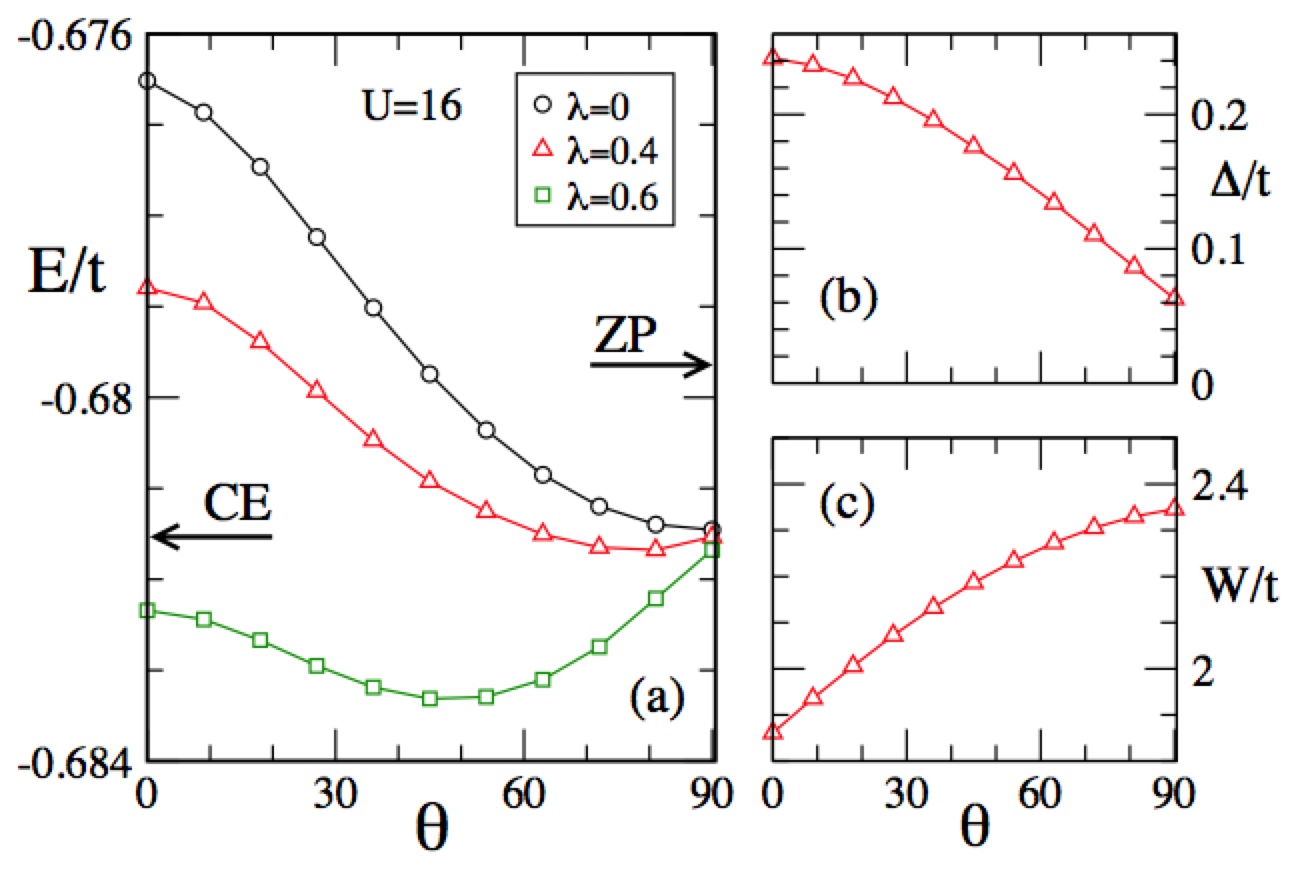}}
\caption{(Color online) Model-hamiltonian results: a) Total energy $E$ vs $\theta$ for different values of  $\lambda$. The $\theta$ dependence of b) DOS-gap $\Delta$, and c) the bandwidth $W$ of the occupied $e_g$.}
\label{fig:3}
\end{figure}

In order to investigate the stability of the CE-type AFM state with respect to the rotation of spin-dimers, we study the degenerate double-exchange model in the presence of the inter-orbital Hubbard repulsion and Jahn-Teller (JT) lattice distortions, with the Hamiltonian,
\begin{eqnarray}
H &=& -\sum_{\langle ij \rangle }^{\alpha \beta}
{t}_{\alpha \beta}^{~ij} \cos(\Theta_{ij}/2)
 c^{\dagger}_{i \alpha } c^{~}_{j \beta } + U \sum_{i} n_{ia}n_{ib}  \cr
&& + J_s\sum_{\langle ij \rangle} {\bf S}_i \cdot {\bf S}_j
 - \lambda \sum_i {\bf Q}_i.{\mbox {\boldmath $\tau$}}_i 
+ {K \over 2} \sum_i {\bf Q}_i^2,
\end{eqnarray}
\noindent
where $c$ ($c^{\dagger}$) is the annihilation (creation) operator, and $\alpha$, $\beta $ are summed over the two Mn-$e_g$ orbitals $d_{x^2-y^2} (a) $ and $d_{3z^2-r^2} (b) $. $t_{\alpha \beta}^{ij}$ denote the nearest-neighbor hopping amplitudes: $t_{a a}^x= t_{a a}^y \equiv t$, $t_{b b}^x= t_{b b}^y \equiv t/3 $, $t_{a b}^x= t_{b a}^x \equiv -t/\sqrt{3} $, $t_{a b}^y= t_{b a}^y \equiv t/\sqrt{3} $. The $\Theta_{ij}$ are the angle between neighboring Mn-$t_{2g}$ spins $\bf{S}_i$ and $\bf{S}_j$ and $J_s$ is their AFM superexchange. $\lambda$ denotes the strength of the JT coupling between the distortion ${\bf Q}_i=(Q_{ix},Q_{iz})$ and the orbital pseudospin ${\tau}^{\mu}_i=\sum_{\alpha\beta} c^{\dagger}_{i\alpha } \Gamma^{\mu}_{\alpha \beta} c_{i\beta }$, where $\Gamma^{\mu}$ are the Pauli matrices. $K$ is a measure of the lattice stiffness (set to unity). Energies are in units of $t$ (estimated  $\sim 0.2 eV$ from the $e_g$ bandwidth in Fig.~\ref{fig:2}a). In the absence of $U$, a CE state with CO/OO is found to be the groundstate over a wide regime in the parameter space~\cite{yunoki}. In the absence of $\lambda$ (and for $U \geq 10$), we find that the $\perp$ state has lower energy than the CE state (see Fig.~\ref{fig:3}a)~\cite{notaenergy}. Inclusion of the JT coupling leads to the stability of a state with an intermediate $ \theta$. Eventually, beyond a ($U$-dependent) critical value of $\lambda$, one recovers the CE state as the groundstate of this model, due to the more pronounced energy gain via JT distortions in the CE state as compared to that in the $\perp$ state.

It is interesting to note that when computing the spin-structure factor for various $\theta$, we find that there is no change in the location or strength of the major peaks, positioned at ($\pi,0$),($0,\pi$) and ($\pi/2,\pi/2$). This makes it very hard to differentiate between the magnetic structures corresponding to different $\theta$ from simple neutron scattering experiments. We also find a decrease in the DOS gap (Fig.~\ref{fig:3}b), and a broadening of the $e_g$ bands (Fig.~\ref{fig:3}c) upon increasing $\theta$, similar to the first-principles results (see Fig.~\ref{fig:2}a). This can be understood in terms of the effective hoppings arising via the DE factor $\cos(\Theta_{ij}/2)$. In the CE state each chain has a perfect hopping, but there is no inter-chain hopping. However, the spin-dimerized state allows for inter-chain hopping at the cost of reduced hopping within chains, leading to an overall gain by a factor $(1+\sqrt2)/2$, which very well explains the increase in the bandwidth $W$ (Fig.~\ref{fig:3}c).

Let us now focus on the multiferroicity. The creation of non-collinear ``spin-dimers'' along with the small but detectable CD breaks IS (present in the CE spin arrangement), allowing FE polarization along the $a$-axis as a realization of the intermediate BC/SC CO-picture proposed by Efremov {\it et al.}~\cite{Efremov04}. There is however one important difference between our work and those model predictions~\cite{Efremov04}: there, the spin-rotation is sufficient to progressively transform a SC-CO into a BC-CO (ZP) and, as such, the $\perp$-structure regains centrosymmetry ({\it i.e.}  leading to the expectation of $P_{ele}$ = 0).  In our case, the spin-rotation -- even for the largest $\theta$ = 90$^o$ -- is not a strong enough factor to produce a BC situation, due to the structural inequivalency between the two kinds of Mn which governs their charge distribution.  Indeed, in our DFT simulations, the larger the deviation from the collinear CE-type, the stronger is the ``asymmetry" introduced by the spin-dimers along the chain and the closer one gets to the ideal realization of the intermediate BC/SC situation, explaining why the maximum is located at $\theta$ = 90$^o$. According to this mechanism, CD is an essential ingredient in the rising of P: there would be no ferroelectricity without CD (despite Mn cations keeping a centrosymmetric distribution). Rather, it is the {\it combination of spin-rotation and CO} that induces ferroelectricity.

\begin{figure}
\resizebox{80mm}{!}{\includegraphics[angle=0]{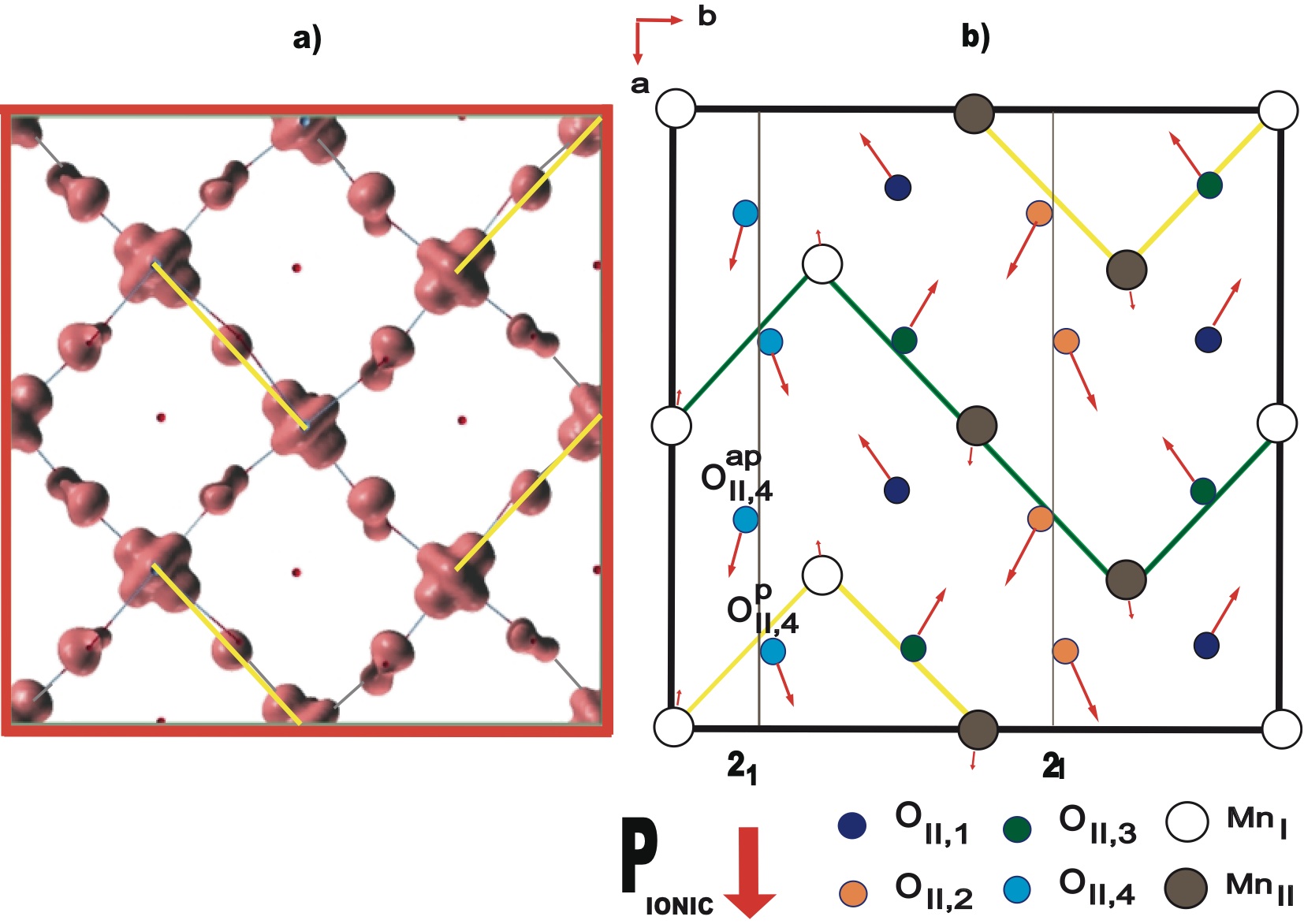}}
\caption{(Color online) a) DFT isosurface of the $e_g$ bands in the LT-O structure. ZP are highlighted. b)
Atomic displacements (not in scale) of the LT-O MnO$_2$ plane. Atoms marked with the same color are structurally equivalent. Zig-zag spin chains highlighted.} 
\label{fig:4}
\end{figure}

We now discuss an alternative lattice structure in which the atomic arrangement is characterized by a {\it structural} Mn-Mn dimerization. In particular, Rodriguez {\it et al.}~\cite{rodriguez} proposed two different LCMO structures (referred in that paper as LT-M and LT-O): the LT-M shows basically the same configuration and symmetries as in Ref.~\cite{radaelli}, whereas the LT-O shows neighboring octahedra in which both Mn are off-centered and with ``long" MnO bonds directed along the same Mn-O-Mn line. This is characteristic of a ZP-like structure~\cite{daoud}, in which the two Mn (despite being still inequivalent by symmetry) are expected to appear electronically more similar than in the SC CO/OO model.  Indeed, when plotting the LT-O isosurface for the $e_g$ manifold (see Fig.~\ref{fig:4}a), the two kinds of Mn show a very similar charge distribution, as also confirmed by the small difference between their moments  (about 3.2 $\mu_B$ and 3.32 $\mu_B$). The peculiar OO shows {\it pairs} of Mn with $3x^2-r^2$ orbitals alternated with pairs of Mn with  $3y^2-r^2$ orbitals, in agreement with the ZP picture. However, from our calculations we do not find an appreciable spin-polarization on the O in the ZP-like dimer,  which is $\sim 0.05 \mu_B$, at variance with much larger values predicted by Hartree-Fock calculations~\cite{patterson,ferrari}. Therefore, small charge-transfer effects are expected, calling for further studies focused on the O spin-polarization (especially in Pr$_{1-x}$Ca$_x$MnO$_3$, $0.3<x<0.5$, where the ZP seems  the ground state). In this regard, we note that with AFM-CE spin configuration the LT-M shows a lower total energy (by $\sim 30$ meV/Mn) than the LT-O, suggesting as the ground state for LCMO a mostly SC state with CD.

The calculation of the FE polarization for this structure in the AFM-CE spin configuration gives particularly interesting results. In the $P2_1nm$ space-group, a non-switchable  polarization is allowed along the $a$-axis~\cite{wu}. However, we here focus on the $b$-axis, investigating the possibility of magnetically switchable multiferroicity.  The $2_1$ symmetry, element of the $P2_1nm$ space-group, forbids any ionic component of $P$ along the $b$-axis: indeed we find $P_{ion}^b=$0. However, when calculating the electronic contribution, a large value is obtained ($P_{ele}^b$ = 7.2 $\mu C/cm^2$), which  should be easily experimentally detected in untwinned high-quality crystals. To investigate the origin of $P$, we show in Fig.~\ref{fig:4}b the displacements of the atoms with respect to a reference centrosymmetric $Pnma$ structure~\cite{radaelli}. It is clear that the components of the displacements along the $a$-axis do not cancel, resulting in a finite $P^a$. However, along $b$, there is a complete cancellation in the displacements, consistent with the $2_1$ symmetry, with the ionic $P$ summing up to zero. Nevertheless, when imposing the AFM-CE spin configuration, the $2_1$ screw axis is not any more a symmetry operation in the {\it magnetic} space-group. For example, the $O$ atoms labeled as $O_{II,4}$, are alternatively bonded to Mn with parallel  ($O_{II,4}^p$) and with antiparallel ($O_{II,4}^{ap}$) spins: they are therefore structurally equivalent but electronically inequivalent, as suggested by the $e_g$ charge density plot (Fig.~\ref{fig:4}a).  To further substantiate the inequivalency of the $O$ as the source for $P_{ele}^b$, we created spin-dimers by coherent rotations. Indeed we find that $P$ decreases continuosly upon increasing $\theta$ and vanishing at 90$^o$ (see Fig.~\ref{fig:1}) as the inequivalence of $O_{II,4}^{p}$ and $O_{II,4}^{ap}$ disappears.

In summary, we have found two different mechanisms for improper ferroelectricity in LCMO. In the first one, starting from centrosymmetric ionic positions and double-zig-zag ferromagnetic spin-chains, an intermediate bond-centered/site-centered polar charge distribution is achieved by means of a spin-dimerization. This induces a dramatic ferroelectric response with $P$ up to few $\mu$C/cm$^2$. The second mechanism, active in the bond-centered ZP-like lattice structure, induces ferroelectricity along $b$-axis, with the AFM-CE spin arrangement lifting the $2_1$ symmetry and paving way to a value of $P_b$ which is largest to date in the whole class of improper magnetic ferroelectrics. Recent electric field gradient experiments on doped manganites have also suggested the existence of ferroelectric domains in the charge ordered regime~\cite{Tokura-EFG}.

Work financially supported by  the European Research Council through the ``BISMUTH" project (Grant N. 203523). 
We thank Daniel Khomskii and Martijn Marsman for valuable discussions.


\begin{thebibliography}{100}
\bibitem{notaenergy} A similar analysis at DFT level should involve a full structural optimization, which is computationally very intensive due to the large unit cell and non-collinear spin states, and is beyond the scope of present work.
\end{thebibliography}
\end{document}